\PassOptionsToPackage{table,dvipsnames}{xcolor}

\documentclass[sigconf]{acmart}
\AtBeginDocument{%
  }

\copyrightyear{2025}
\acmYear{2025}
\setcopyright{acmlicensed}\acmConference[CUI '25]{Proceedings of the 7th ACM Conference on Conversational User Interfaces}{July 8--10, 2025}{Waterloo, ON, Canada}
\acmBooktitle{Proceedings of the 7th ACM Conference on Conversational User Interfaces (CUI '25), July 8--10, 2025, Waterloo, ON, Canada}
\acmDOI{10.1145/3719160.3736616}
\acmISBN{979-8-4007-1527-3/2025/07}

\usepackage{graphicx}
\usepackage{subcaption}
\usepackage{multirow}
\usepackage{amsmath}
\usepackage{array}
\usepackage{stfloats}

\begin{document}


\title{Visual-Conversational Interface for Evidence-Based Explanation of Diabetes Risk Prediction}


\author{Reza Samimi}
\orcid{0000-0001-6751-1559}
\affiliation{%
  \institution{KU Leuven}
  \city{Leuven}
  \country{Belgium}}
\email{reza.samimi@kuleuven.be}

\author{Aditya Bhattacharya}
\orcid{0000-0003-2740-039X}
\affiliation{%
  \institution{KU Leuven}
  \city{Leuven}
  \country{Belgium}}
\email{aditya.bhattacharya@kuleuven.be}

\author{Lucija Gosak}
\orcid{0000-0002-8742-6594}
\affiliation{%
  \institution{University of Maribor}
  \city{Maribor}
  \country{Slovenia}}
\email{lucija.gosak2@um.si}

\author{Gregor Stiglic}
\orcid{0000-0002-0183-8679}
\affiliation{%
  \institution{University of Maribor}
  \city{Maribor}
  \country{Slovenia}}
\email{gregor.stiglic@um.si}

\author{Katrien Verbert}
\orcid{0000-0001-6699-7710}
\affiliation{%
  \institution{KU Leuven}
  \city{Leuven}
  \country{Belgium}}
\email{katrien.verbert@kuleuven.be}

\renewcommand{\shortauthors}{Samimi et al.}

\begin{abstract}
Healthcare professionals need effective ways to use, understand, and validate AI-driven clinical decision support systems. Existing systems face two key limitations: complex visualizations and a lack of grounding in scientific evidence. We present an integrated decision support system that combines interactive visualizations with a conversational agent to explain diabetes risk assessments. We propose a hybrid prompt handling approach combining fine-tuned language models for analytical queries with general Large Language Models (LLMs) for broader medical questions, a methodology for grounding AI explanations in scientific evidence, and a feature range analysis technique to support deeper understanding of feature contributions. We conducted a mixed-methods study with 30 healthcare professionals and found that the conversational interactions helped healthcare professionals build a clear understanding of model assessments, while the integration of scientific evidence calibrated trust in the system's decisions. Most participants reported that the system supported both patient risk evaluation and recommendation.
\end{abstract}





\maketitle

\section{Introduction}
The integration of artificial intelligence (AI) into healthcare decision support systems has shown promising potential for improving patient care and clinical outcomes \cite{Reddy2021, Secinaro2021, Alowais2023}. However, the successful adoption of AI systems in healthcare depends on healthcare professionals' (HCPs) ability to understand, appropriately trust, and effectively utilize AI-generated recommendations \cite{Shinners2019, Asan2020}. This is particularly crucial in medical decisions, where AI systems are increasingly being employed to assist in diagnosis, risk assessment, and treatment planning \cite{Rony2024, Kerstan2023}.

Clinical Decision Support Systems (CDSS) are software applications that assist healthcare professionals in clinical decision-making by providing patient-specific assessments and recommendations. These systems analyze patient data through rule-based algorithms, statistical models, or machine learning approaches to generate insights for improving care \cite{sutton2020overview}. The healthcare professional's workflow with a CDSS encompasses several integrated steps : patient data is captured from electronic records or direct entry; the system processes this information through analytical algorithms to generate clinical insights; these insights are communicated via intuitive visualizations or summaries; clinicians then evaluate these recommendations by comparing them with established medical knowledge and their own expertise; and finally, they translate validated insights into specific clinical actions like diagnostic procedures or treatment adjustments \cite{foraker2015ehr, sutton2020overview}.

A barrier to the widespread adoption of AI in clinical practice through decision support systems is the inherent complexity and opacity of AI decision-making processes \cite{hassan2024barriers}. In recent years, significant advances have been made in explainable AI (XAI) to help users understand complex models in healthcare \cite{loh2022application, rajashekar2024human}. An extensive set of visualizations has been developed to represent complex models and their features, especially for non-expert users who do not have the required technical AI knowledge to understand the working of such black box models\cite{cheng2019explaining} \cite{ooge2022explaining}; however, many of these require significant expertise to interpret. 

Recent work in the XAI domain has highlighted the promising role of conversational approaches in explaining AI decisions \cite{nguyen2022explaining, nobani2021towards}. Through natural language interaction, these approaches enable non-experts in AI to ask questions about specific aspects of model predictions, receive immediate clarifications, and explore model reasoning through an intuitive dialogue format that mirrors their clinical case discussions. This interactive question-answering process has proven more accessible than static visualizations, as it allows users to progressively build understanding through targeted follow-up questions about aspects most relevant to their decision-making \cite{Slack2023}. 

It has been argued that simply explaining a single decision without the origins of that action or its context will not carry sufficient meaning \cite{dazeley2021levels}. Traditional approaches to building appropriate trust in AI-powered clinical decision support systems have primarily relied on explaining the AI's decision-making process and performance metrics \cite{rechkemmer2022confidence, wang2019designing}. However, research has shown that these explanations alone often fail to persuade clinicians to accept correct AI recommendations when they conflict with clinicians' initial hypotheses, and may not help clinicians identify incorrect AI suggestions \cite{yang2019unremarkable, holzinger2017towards, xie2020chexplain}.

Recent studies have demonstrated a more effective approach: integrating scientific evidence alongside AI predictions. This evidence-based approach, which aligns with clinicians' natural suggestion-validation processes, has shown promise in helping healthcare professionals appropriately calibrate their trust in AI systems \cite{yang2023harnessing}. Furthermore, research indicates that scientific evidence of the impact of predictive DSS on patient care plays an important role in facilitating trust \cite{schwartz2022factors}. However, providing only scientific evidence is insufficient, as it does not reflect how machine learning models actually arrive at their decisions. Healthcare professionals require explanations into model decison-making to judge whether the system’s output
is trustworthy \cite{tonekaboni2019clinicians}.

Our research focuses on bridging this gap between AI systems and evidence-based clinical practice and supporting healthcare professionals in understanding and validating AI-driven clinical decision support systems. This context allows us to examine how integrated visual and conversational explanations and the inclusion of scientific evidence can support clinical decision-making through three key research questions:

\begin{enumerate}
    \item How do conversational explanations impact the understandability of HCPs in AI's decision-making?
    \item How does the inclusion of scientific evidence in explaining AI's decision-making calibrate the trust of HCPs in the AI's decisions? 
    \item How do HCPs find our integrated visual and conversational explanations useful in evaluating the patient’s risk of diabetes as well as providing recommendations to improve patient conditions?
\end{enumerate}

The main objective of this research is to determine whether conversational explanations can enhance the understandability of AI decision-making processes for non-expert users, with particular emphasis on making visual explanations more comprehensible for healthcare professionals. To explore the research questions, we conducted a mixed-methods study with 30 healthcare professionals. The study utilized both quantitative measures and qualitative feedback to evaluate our system's effectiveness in terms of understandability, usefulness, actionability, and trust. Prior to the main study, we conducted multiple co-design sessions with nurses specializing in diabetes care and bioinformatics to refine the system's design and functionality. These sessions helped shape key features of the system, particularly its practical utility and the development of actionable recommendations. 

Our mixed-methods study with healthcare professionals revealed that the conversational interface supported understanding of AI assessments, with queries primarily focusing on model explanations. Scientific evidence integration positively influenced trust calibration while encouraging critical evaluation rather than unconditional acceptance. Most participants actively utilized both visual and conversational components for risk analysis and recommendation development, though they identified needs for greater personalization and more comprehensive patient data. 

This paper makes the following contributions:
\begin{enumerate}
\item As an \textit{artifact contribution}, we present an decision support system with a novel design that integrates interactive visualizations with conversational AI for diabetes risk assessment, enabling healthcare professionals to efficiently explore AI assessments while maintaining structured analysis capabilities. The code is open-sourced on {GitHub}\footnote{\url{https://github.com/resamimi/diabetes-dashboard/}}.

\item As \textit{theoretical contributions}, we introduce: (1) an approach for grounding AI explanations in scientific evidence, enabling HCPs to validate AI findings against established medical knowledge and bridging the gap between AI systems and evidence-based clinical practice; (2) a hybrid approach for handling user prompts that combines specialized models for analytical functions with general LLMs for broader queries, ensuring support for healthcare professionals' diverse information needs; and (3) a feature range analysis technique that identifies "AI-observed ranges" of values most influential in predictions, enabling systematic examination of how AI systems utilize different value ranges and providing deeper insight on feature contribution in decision-making.

\item As an \textit{empirical contribution}, we present findings from a mixed-methods study with 30 healthcare professionals and multiple co-design sessions, revealing that healthcare professionals build appropriate trust incrementally through evidence-grounded explanations rather than accepting AI recommendations at face value, and integrated visual and conversational interfaces enhance clinical decision-making by supporting both analytical reasoning and contextual understanding.
\end{enumerate}

\section{Related Work and Background}\label{sec:background}
This section reviews relevant advances and remaining challenges in explainable AI approaches in clinical DSSs, conversational explanations and interfaces, and the integration of evidence-based medicine with AI systems.

\subsection{Clinical Decision Support Systems with Explainable AI}
The adoption of AI in healthcare has brought advances in diagnosis, prognosis, and treatment planning \cite{Bharati_2024}, but the complexity of AI systems challenges healthcare professionals who need to understand AI-generated recommendations \cite{amann2020explainability}. Traditional XAI methods in healthcare have focused on visualization techniques such as feature importance plots and saliency maps \cite{sadeghi2023brief}, which often require technical expertise to interpret effectively \cite{rober2024clinicians}.

Decision Support Systems (DSS) in healthcare face challenges in balancing functionality with usability in clinical environments \cite{kushniruk2011issues, o2014decision}. Visualization-based interfaces have attempted to present complex medical data in more intuitive formats \cite{bhattacharya2023directive,kwon2018retainvis,rostamzadeh2021visual}, but healthcare professionals still face challenges in translating visual patterns into actionable clinical insights.

Recent XAI developments have introduced more sophisticated approaches, including counterfactual explanations and example-based reasoning \cite{dai2022counterfactual} \cite{mindlin2025beyond}. Conversational XAI represents a shift toward accessibility, with systems like GutGPT \cite{rajashekar2024human} demonstrating how LLM-augmented interfaces can mirror clinical consultation processes by enabling domain-specific dialogue with evidence-grounded responses, though participants often found visualization components like Partial Dependency Plots and Individual Conditional Expectation plots challenging to interpret.

\subsection{Conversational Explanations and Interfaces}
Recent advances in Large Language Models (LLMs) have enabled more sophisticated approaches to AI explanations through natural dialogue \cite{Slack2023, wang2024llmcheckup}. These models make complex AI decisions more accessible through open-ended conversations, though most focus solely on generating natural language responses without integrating visual elements or maintaining consistent grounding in domain expertise \cite{feldhus2023interrolang}.

In healthcare, conversational interfaces have been explored for patient engagement, clinical decision support, and medical education \cite{wen2024leveraging}. Healthcare professionals appreciate natural dialogue interfaces that mirror clinical conversations \cite{rajashekar2024human, laranjo2018conversational}, but existing healthcare chatbots typically operate in isolation from clinical tools and visualization systems \cite{xing2019conversational} and rarely connect medical knowledge bases to specific patterns learned by AI models \cite{alowais2023revolutionizing}.

Prior research has demonstrated that visualizations offer distinct advantages in healthcare contexts, facilitating quick understanding of patient information \cite{bhattacharya2023directive}. Concurrently, research shows that textual explanations can enhance interpretability of complex visualizations, as healthcare professionals often find technical visualizations difficult to interpret without accompanying textual context \cite{bhattacharya2023directive, rajashekar2024human}. However, existing systems often fail to maintain consistent grounding in scientific evidence, potentially reducing trust from healthcare practitioners \cite{kaufman2023explainable}.

\subsection{Evidence-based Medicine and AI Integration}
Evidence-based medicine (EBM) forms the foundation of modern clinical practice, requiring healthcare professionals to integrate clinical expertise with the best available external clinical evidence \cite{sackett1996evidence}. A challenge lies in bridging the gap between statistical patterns identified by machine learning models and causal relationships established through clinical research \cite{gibney2024has}.

Current approaches to incorporating evidence into AI systems typically either use scientific literature to inform model development or attempt to justify AI predictions through post-hoc evidence retrieval \cite{wang2024accelerating, patel2024retrieve}. Recent work has explored incorporating scientific evidence into AI explanations \cite{procter2023holding}, including knowledge graphs connecting AI predictions with relevant clinical studies \cite{patel2024retrieve}.

A promising approach integrates scientific evidence alongside AI predictions, aligning with clinicians' natural suggestion-validation processes \cite{yang2023harnessing}. This helps healthcare professionals calibrate trust by providing literature evidence that can validate or invalidate AI suggestions. Similarly, GutGPT \cite{rajashekar2024human} showed that LLM-augmented systems with guideline citations improved trust. However, these approaches either don't explain the AI's decision-making process, which healthcare professionals require to evaluate AI trustworthiness \cite{tonekaboni2019clinicians}, or lack mechanisms for clinicians to systematically compare AI explanations against established medical knowledge.

\subsection{Clinical Decision Support Systems for Type 2 Diabetes}
Recent years have seen numerous clinical decision support systems (CDSS) for Type~2 Diabetes (T2D), providing interactive visualizations, risk assessment models, EHR integration, patient stratification, and personalized treatment recommendations.

Bhattacharya et al.~\cite{bhattacharya2023directive} present an explainable CDSS that predicts T2D risk via an interactive dashboard with visual explanations and "what-if" scenarios. Kent et al.~\cite{kent2022ehr} developed an EHR-integrated risk stratification tool, while Kourou et al.~\cite{kourou2021integration} found that standalone CDSS are less useful when not integrated with EHR systems. Other approaches include OnT2D-DSS~\cite{spoladore2024ont2d}, an ontology-driven system for tailored nutrition therapy plans, and \textit{Exandra}\cite{grechuta2025exandra}, which recommends evidence-based therapies aligned with diabetes guidelines.

Existing T2D systems have limitations: they either rely on visualizations requiring specific interpretation skills without sufficient explanation, or they fail to provide explanations grounded in scientific evidence, challenging healthcare professionals' ability to trust and validate recommendations.
\\
\\
Our work addresses gaps in existing systems by integrating visual explanations with conversational capabilities. Building on previous visualization designs that have proven effective in facilitating quick understanding of healthcare professionals for data-centric and feature importance explanations \cite{bhattacharya2023directive}, we improve these with an interactive conversational interface that enables users to ask follow-up questions about specific visual elements to clarify any confusion or get more information. This integration proves particularly helpful for healthcare professionals by allowing them to quickly gain a clear understanding of consolidated patient information and factors contributing to risk through visualizations while accessing deeper contextual insights through dialogue. 

Moreover, the system bases AI explanations on scientific evidence, drawing on verified evidence from medical journals, clinical guidelines, systematic reviews, and epidemiological studies, enabling direct comparison between AI predictions and established medical knowledge. This is achieved through our feature range analysis visualization that explicitly juxtaposes AI-observed ranges with scientifically established ranges from medical literature, allowing healthcare professionals to directly evaluate how well the AI's learned patterns align with clinical understanding.

\subsection{Technical Background}
To explain our approach, we present key technical foundations in two critical areas: natural language processing approaches for model explanations, and methods for analyzing and validating feature importance in AI models.

\subsubsection{Natural Language Processing for Model Explanations} \label{sec:nlp_processing}
A key challenge in making AI models interpretable through natural language interaction is handling diverse user queries about model behavior. The TalkToModel \cite{Slack2023} framework addressed this by using pre-trained language models to parse user prompts and map them to specific backend functionalities such as feature importance analysis and counterfactual explanations. Their approach employed few-shot GPT-J and fine-tuned T5 models using a dataset of prompts paired with corresponding parsed versions that mapped to specific backend operations.

However, this approach has inherent limitations. The models could at best handle prompts similar to those in their training dataset. When presented with unfamiliar queries, they either produce irrelevant responses or default to "I did not understand your question." These unsupported queries fell into two categories: those with meanings entirely different from any supported functionality, and those asking about supported functionalities but using unfamiliar phrasings. These limitations are addressed in our system (described in Section \ref{sec:implementation}).

\subsubsection{Feature Importance Analysis Methods} \label{sec:TTM_feature_importance}
Various post hoc explanation methods exist, each providing feature attributions through different computational approaches: Local Interpretable Model-Agnostic Explanations (LIME) \cite{ribeiro2016should} fits local linear models, Shapley Additive Explanations (SHAP) \cite{lundberg2017unified} estimates Shapley values, and Integrated Gradients \cite{sundararajan2017axiomatic} utilize model gradients.

To assess explanation reliability, the most significant features are perturbed to measure impact on model predictions. The fudge score quantifies prediction sensitivity to feature perturbations:

\begin{equation}
\text{Fudge}_\mathcal{N}(f, x, m) = \frac{1}{N} \sum_{n=1}^{N} \left| f_\mathcal{N}(x) - f_\mathcal{N}(x + \epsilon_n \odot m) \right|
\end{equation}

where \( \odot \) represents element-wise product, and \( \epsilon \sim \mathcal{N}(0, I\sigma) \) is Gaussian noise. To quantify explanation faithfulness, we compute the area under the fudge score curve for the top-\( k \) most important features:

\begin{equation}
\mathcal{N}(k, \phi) = \begin{cases} 1, & \phi_i \in \arg \max\limits_{\phi' \subset \{1...d\}, |\phi'|=k} \sum_{i \in \phi'} |\phi_i| \\
0, & \text{otherwise} \end{cases}
\end{equation}

\begin{equation}
\text{Faith}_\mathcal{N}(\phi, f, x, K) = \sum_{k=1}^{K} \text{Fudge}_\mathcal{N}(f, x, \mathcal{N}(k, \phi))
\end{equation}

where \( \mathcal{N}(k, \phi) \) selects the top-\( k \) features. Higher faithfulness scores indicate correctly identified crucial features whose perturbation affects predictions.

To select the optimal explanation, we compute faithfulness scores for multiple methods (LIME with kernel widths \{0.25, 0.50, 0.75, 1.0\} and KernelSHAP with default settings) and choose the highest scoring method. We set \( \sigma = 0.05 \) to maintain local perturbations and \( K = \lfloor d/2 \rfloor \). For similar scores (within \( \delta = 0.01 \)), we use Jaccard similarity of feature rankings as a tiebreaker instead of the \( L_2 \) norm to ensure stability across methods.

\section{Technical Implementation} \label{sec:implementation}
This section provides in-depth details about the implementation and design process of our system for healthcare decision support. While our approach is designed to be adaptable to various clinical contexts, we specifically tested and applied it in the domain of type 2 diabetes risk prediction.

\subsection{System Architecture and Components}
\begin{figure*}[t!]
  \begin{subfigure}{\textwidth}
    \includegraphics[width=\textwidth]{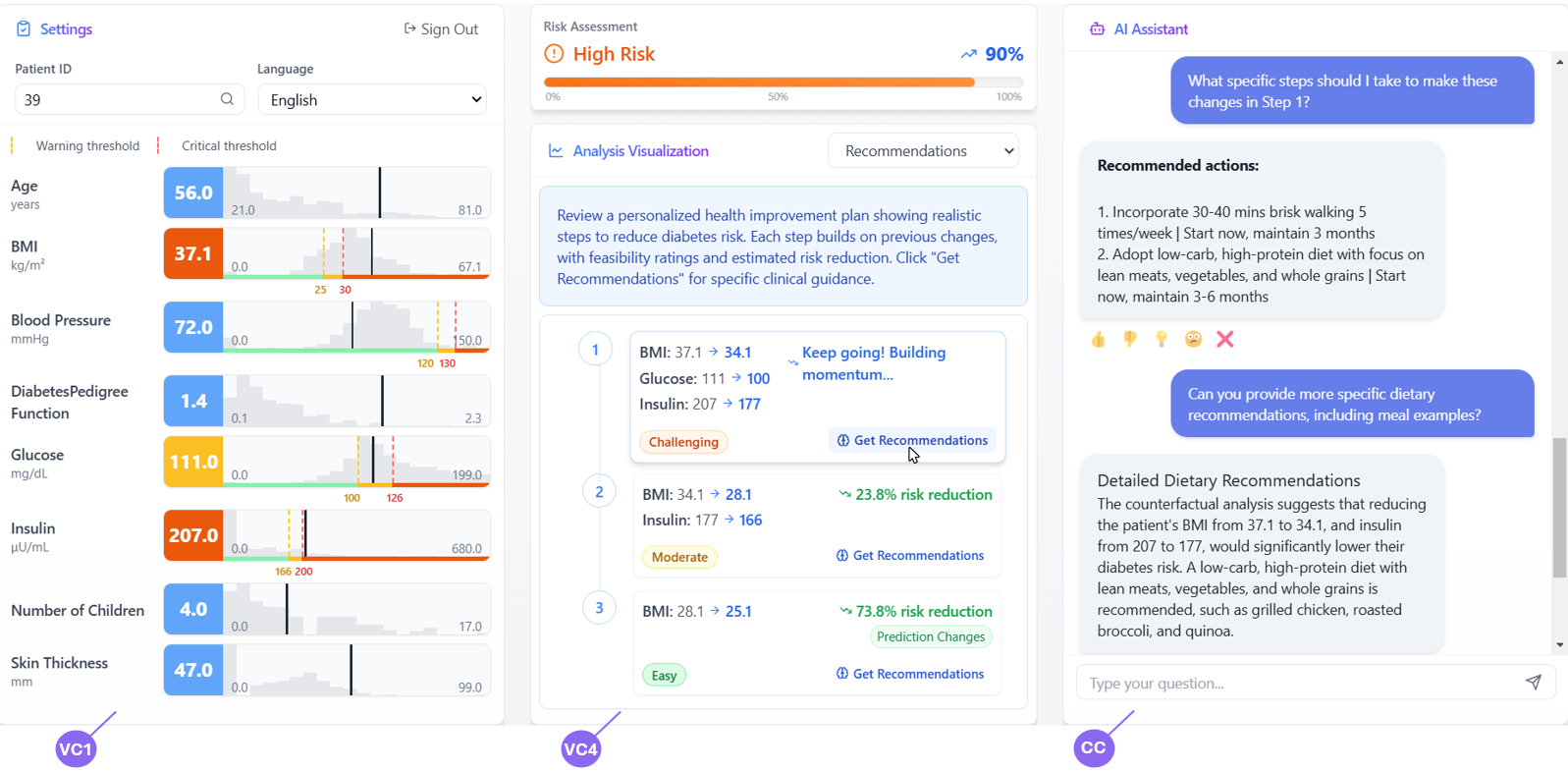}
    \caption{The overall interface of our DSS: the patient record data (VC1) is visualized in the left column. In the middle, The user can see the risk assessment of the chosen patient and select different visualizations from the drop-down menu of Analysis Visualization (in this case Recommendations (VC4)). The chat component (CC) in the right column provides responses to the user interactions (such as clicking on "Get Recommendation" button) as well as follow-up questions.}
    \label{fig:dashboard}
  \end{subfigure}
  
  \vspace{1em}
  
  \begin{subfigure}{\textwidth}
    \includegraphics[width=\textwidth]{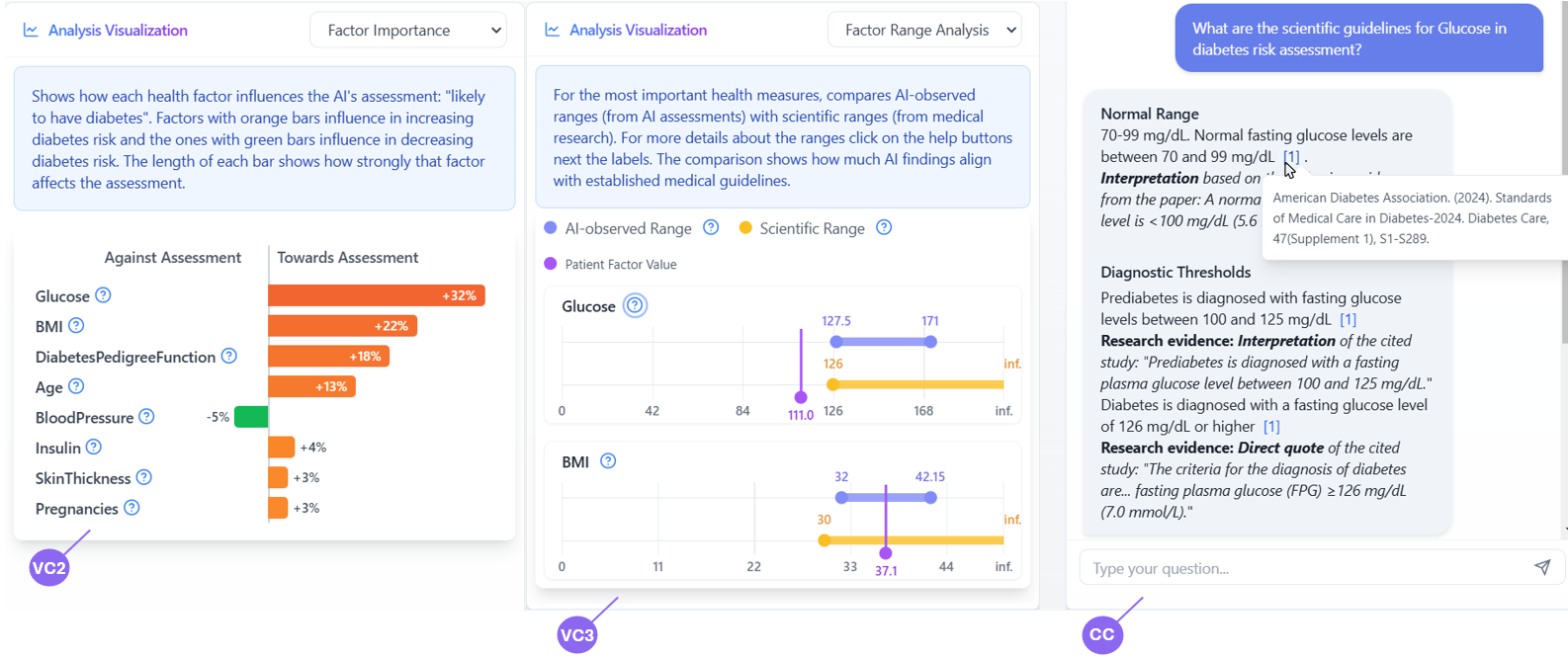}
    \caption{The left and middle visualizations correspond to feature importance (VC2) and feature range (VC3) analysis, respectively. When the user clicks on the help buttons next to the feature names, the scientific evidence supporting the provided importance and range appears in the chat area. In this case, the evidence for the normal and diagnostic ranges is shown. The user can also see references for the evidence by hovering over the in-text citations.}
    \label{fig:factor_contributions}
  \end{subfigure}
  
  \caption{The DSS interface components. (a) Shows the main dashboard with patient records, analysis visualizations, and chat component. (b) Displays feature importance and range analysis with scientific evidence integration.}
  \Description{Two screenshots of the decision support system interface. The top image shows the complete dashboard with three main components: patient record data visualization on the left, risk assessment and analysis visualization in the middle, and a chat component on the right. The bottom image shows feature importance analysis and feature range analysis visualizations with scientific evidence information displayed in the chat area when users interact with help buttons.}
  \label{fig:system_interface}
\end{figure*}

In this part, the implementation details of different components of the system are discussed. The interface of the DSS is shown in Figure \ref{fig:dashboard}.

\begin{figure}[t!]
  \includegraphics[width=0.5\textwidth]{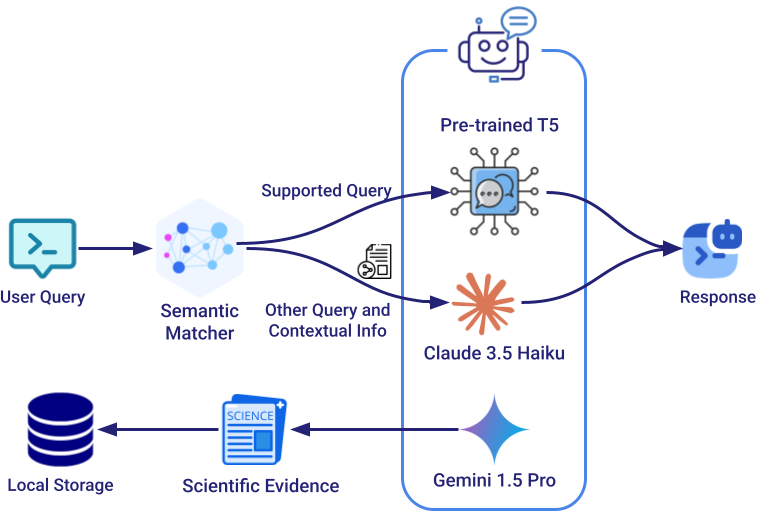}
  \caption{Hybrid query processing architecture for conversational AI integration. Incoming user queries are first processed by a semantic matcher that determines if they match supported analytical operations. Matching queries are parsed by a fine-tuned T5 model to trigger specific backend functions, while non-matching queries are handled by a general-purpose LLM (Claude) that has access to contextual information. The scientific evidence are retrieved by Gemini}
  \Description{A diagram illustrating the hybrid query processing architecture of the chatbot. The workflow shows user queries being evaluated by a semantic matching component that routes them either to a specialized T5 model for analytical functions (if similar to supported operations) or to Claude LLM for general healthcare questions. The system provides contextual information to Claude, including conversation history, patient data, and current visualization content to ensure relevant and accurate responses.}
  \label{fig:ai_assistant}
\end{figure}

\subsubsection{Chatbot Implementation} 

Our system builds upon the TalkToModel framework (Section \ref{sec:nlp_processing}) while introducing several enhancements to improve robustness and coverage. We developed a hybrid approach that using semantic matching with a specialized sentence transformers model, named all-MiniLM-L6-v2, to ensure comprehensive handling of user queries. The hybrid approach is shown in Figure \ref{fig:ai_assistant}.

\paragraph{Hybrid Query Processing Architecture}
The semantic matcher model is specifically fine-tuned on the user prompts from the TalkToModel \cite{Slack2023} dataset\footnote{The dataset is made available in the supplementary materials}. This fine-tuning process enables the model to accurately determine whether incoming queries are semantically similar to supported operations such as feature importance, counterfactual explanations, etc. When a new query is received, the system compares its semantic representation to those of known supported prompts, using a carefully calibrated similarity threshold to determine if the query maps to existing functionality. For queries that match supported operations, we utilize the existing fine-tuned T5 small model to parse the input into formal grammar patterns that trigger specific backend functionalities. Queries that fall below the similarity threshold for supported operations are redirected to Claude Haiku via API, ensuring the system can still provide meaningful responses to a broader range of questions while maintaining high accuracy for core analytical functions. An example of a follow-up question that is not among the initial supported prompts is shown in Figure \ref{fig:dashboard}.  

\paragraph{Model Selection Rationale}
The fine-tuned T5 small model was chosen over the few-shot GPT-J models and the T5 large model because, first, as demonstrated in the TalkToModel document, fine-tuned T5 models outperformed few-shot GPT-J models in accuracy. Second, while testing the T5 large model, we found that it introduced noticeable latency in response times without providing better handling of queries that diverged from supported prompts.

While more recent Small Language Models (SLMs) like Phi-3, Llama-7B, or other alternatives could potentially be utilized, we leveraged the pre-trained models already made available by TalkToModel on this specific dataset to ensure reproducibility. Furthermore, it's important to note that regardless of which SLM is fine-tuned for specialized analytical tasks, it would still need to be coupled with a general-purpose LLM (like Claude in our implementation) to provide appropriate responses to the broad range of healthcare questions that fall outside the scope of model-specific analytical queries.

\paragraph{Contextual Grounding Mechanism}
To ensure accurate and contextually grounded responses from Claude, the system provides it with comprehensive contextual information including the user's last three conversations with the chatbot (if any), the current patient's health factor values, and the data being displayed in the active Analysis visualization including the scientific evidences supporting the explanations. Claude was selected among high-performing language models because it allows transmission of a large volume of contextual information and manages large data effectively \cite{cao2024characterizing} which helps prevent hallucination as much as possible and enables more precise, situation-aware responses.

\paragraph{Scientific Evidence Integration}
To ground the system's explanations in established medical knowledge, we initially leveraged Gemini 1.5 Pro to gather scientific evidence about diabetes risk factors, their importance in diagnosis, and their clinical ranges. This process involved extracting information from multiple authoritative sources, including peer-reviewed research papers from medical journals, clinical practice guidelines from organizations like WHO\footnote{\url{https://www.who.int/health-topics/diabetes}} and ADA\footnote{\url{https://diabetes.org/}}, systematic reviews and meta-analyses, and large-scale epidemiological studies. Importantly, to ensure the validity and reliability of this evidence base, all AI-retrieved references were manually verified and stored in a dedicated knowledge base. This approach means that when users request scientific evidence for a particular factor or range, the system draws from this pre-verified repository rather than generating new AI responses, ensuring consistency and accuracy in the scientific backing for its explanations. An example of the evidence that the system provides is demonstrated in Figure \ref{fig:factor_contributions}.

\subsubsection{Feature Importance Visualization}

The feature importance analysis uses the computational approach established in TalkToModel \cite{Slack2023} (Section \ref{sec:TTM_feature_importance}) to calculate the importance (contribution) of each feature in the model prediction. For visualization of these values, we adopted the design principles from Bhattacharya et al. \cite{bhattacharya2023directive}, which was shown to be intuitive for healthcare professionals in presenting feature contributions. The importance of each feature is visualized in the DSS interface (Figure \ref{fig:factor_contributions}). Users can access scientific evidence extracted from established medical research papers on each factor's importance by clicking help buttons placed next to each factor.

\subsubsection{Feature Range Analysis}

We propose analyzing factor ranges to give more in-depth insight about the influence of the features in the machine learning model's decision-making in diabetes predictions. Specifically, this analysis identifies ranges of the most important factors (previously determined) that contribute most to the model's predictions. This approach is based on the fact that machine learning models make predictions based on learned patterns from data.

To determine the most contributing ranges, we follow this process for each prediction class (e.g., likely to have diabetes):

\begin{enumerate}
    \item Filter all samples in the dataset predicted in the same class
    \item Identify the 25th and 75th percentiles of values for each important factor
    \item Define the range between these percentiles as the source of the factor's influence on predictions
\end{enumerate}

While we experimented with broader ranges (10th to 90th percentiles), we observed that the 25-75 percentile range provided optimal results, representing the largest range that aligns well with established medical ranges while minimizing outlier influence. This range encompasses values from 50 percent of similarly predicted samples.

The visualization compares these "AI-observed ranges" with scientific ranges established from medical papers and clinical guidelines. This comparison enables healthcare professionals to validate the ranges against their clinical knowledge. Users can access the scientific evidence for the extracted scientific ranges through help buttons placed next to each factor (Figure \ref{fig:factor_contributions}).

\subsubsection{Patient Record Visualization}
The patient record visualization was designed to provide a comprehensive yet intuitive display of health factors. The interface presents each health metric in a format that supports both quick assessment and detailed analysis through several carefully chosen design elements:

\begin{itemize}
    \item \textit{Distribution Context:} Building on established visualization patterns in healthcare interfaces \cite{bhattacharya2023directive}, each metric is shown against a background distribution graph rendered in grey, providing immediate context for how individual values compare to the broader patient population. This design choice helps healthcare professionals quickly identify unusual or concerning values.
    
    \item \textit{Range Markers:} we incorporate minimum and maximum value indicators from the dataset to establish clear boundaries for normal ranges.
    
    \item \textit{Clinical Thresholds:} We implemented a dual-threshold system for critical health indicators, using color-coding and visual markers to highlight warning and critical levels. This design ensures immediate visibility of concerning values while maintaining a clear distinction between different risk levels.
\end{itemize}

The thresholding system was specifically implemented for four key factors based on established clinical guidelines:
\begin{itemize}
    \item BMI: Overweight and obesity status
    \item Blood Pressure: Pre-hypertension and hypertension
    \item Glucose: Pre-diabetes and diabetes
    \item Insulin: Potential and actual insulin resistance
\end{itemize}

Following previous research highlighting the importance of deeper inspection for "actionable" health factors \cite{bhattacharya2024exmos}, this targeted thresholding approach was selectively applied to these modifiable factors that clinicians can directly address through interventions. This visualization design prioritizes rapid clinical assessment while maintaining access to detailed information, supporting healthcare professionals in making informed decisions about patient care.

\subsubsection{Recommendation System}
Our recommendation system enhances the counterfactual analysis from TalkToModel by addressing key limitations in the original implementation. While TalkToModel provided options for prediction changes (such as reducing BMI from 37 to 21), these options were often unrealistic for real patients to implement. Additionally, some options included impossible changes to immutable factors like Age \cite{bhattacharya2024exmos}.

Our system implements several improvements to ensure practical utility. First, it systematically filters out options that include impossible changes, such as modifications to immutable patient characteristics. Following established approaches in medical visualization \cite{bhattacharya2023directive}, the system breaks down unrealistic large-scale modifications into smaller, achievable steps that patients can reasonably implement. Each step is evaluated against medical guidelines to assign a feasibility badge, providing healthcare professionals with clear indicators of the difficulty and practicality of suggested changes.

We introduce two features to enhance the recommendation process. First, we implement a sequential timeline visualization that shows progressive steps until the prediction changes, helping healthcare professionals understand the cumulative impact of incremental improvements. Second, we integrate a "Get Recommendation" button that triggers the chatbot to generate precise, time-bound actions for each step. This allows users to ask follow-up questions about to get more detailed recommendations.

The final recommendations combine insights from multiple sources:
\begin{itemize}
    \item Counterfactual analysis to identify mathematically valid changes
    \item Clinical guidelines to ensure medical appropriateness
    \item Sequential timeline analysis to support gradual implementation
    \item Natural language generation for actionable guidance
\end{itemize}

Figure \ref{fig:dashboard} demonstrates this integrated approach, showing both the step-by-step visualization and an example of the chatbot's detailed recommendations.

\subsection{Model and Dataset}
The system employs a gradient boosted tree model for binary classification of diabetes risk prediction. We utilize the Pima Indians Diabetes Database from the UCI Repository of Machine Learning Databases (now available through Kaggle) \cite{pima_indians_diabetes}. This dataset is widely used in the machine learning community \cite{chang2023pima}, including in intelligent interface research \cite{gomez2020vice}, making it a suitable choice for evaluating explanation approaches.

The dataset contains 768 samples with features including glucose levels, blood pressure, BMI, and other relevant health metrics. The model achieves 73.3\% accuracy on the test set which includes 40\% of all data. While more complex algorithms might potentially achieve higher accuracy, this performance level is sufficient for our research objectives focusing on explanation methods and interface design rather than prediction optimization. 

Our system architecture is designed with model-agnostic principles, allowing any machine learning model trained on the dataset to be easily incorporated without requiring changes to the interface layout or explanation framework. While the choice of model does not influence the system's design and visualization capabilities, the dataset characteristics directly impact visualization components as they must be adaptable to the specific features present in the dataset, their ranges, and distributions.
\subsection{User-Centered Design Process}
Prior to implementing the system, we conducted two co-design sessions with two female nurses between 20-30 years old, knowledgeable in diabetes diagnosis and bioinformatics. These sessions were organized and facilitated by the principal researcher and a supervising professor with expertise in human-computer interaction. Both sessions were conducted online, lasting approximately one and a half hours combined.  The co-design sessions were recorded and transcribed; then, the key points of their conversations were extracted to inform the next iteration of the design. Analysis focused on identifying specific pain points, user preferences, and suggestions for improvement. 

\begin{figure*}[t!]
  \includegraphics[width=\textwidth]{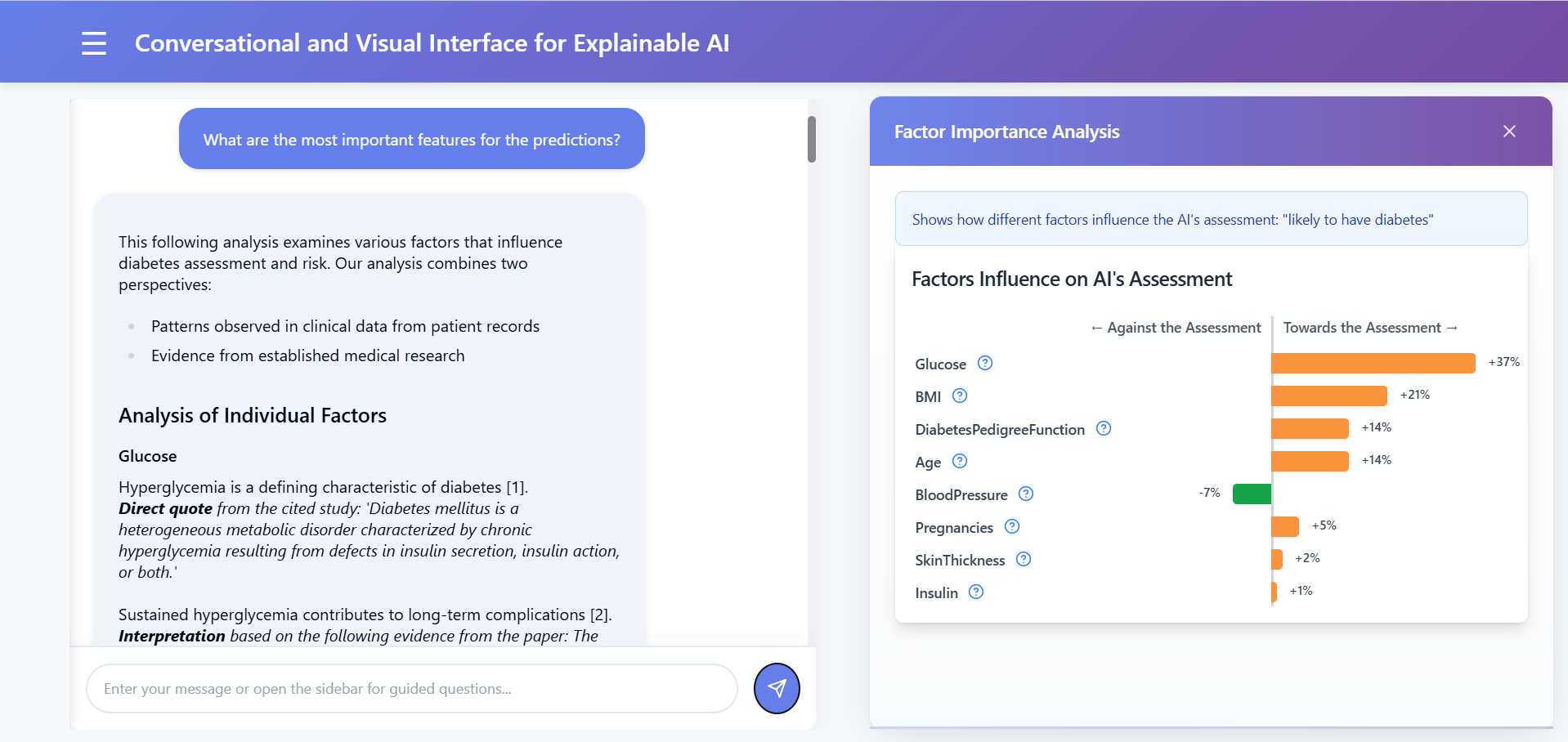}
  \caption{The initial prototype featuring a conversational UI with a chat interface and visualization canvas.}
  \label{fig:chat_ui}
  \Description{A screenshot of the initial prototype showing an interface with a chat area on the left side where users can ask questions about explanations, and a visualization canvas on the right side that displays the corresponding visual representation when requested through the chat. The interface demonstrates how explanations could be accessed through conversation while providing visual context.}
\end{figure*}

Initially, our approach to feature importance and range explanations took a global perspective, implementing these analyses across the entire dataset rather than for individual samples. This strategy focused on providing ``broad'' explanations of general model behavior, drawing insights from multiple algorithmic approaches to offer a comprehensive view of the system's decision-making patterns. The counterfactual explanations in this prototype only provided mathematical options for changing health factor values to alter a prediction, without practical implementation guidance. In the first co-design session, we presented an initial prototype that featured a primarily conversational UI (shown in Figure \ref{fig:chat_ui}). The prototype employed a chat interface where, when users asked for explanations (e.g., feature importance), the corresponding visualization would appear in a canvas displayed adjacent to the chat area. Participants were shown several explanation methods, including patient record, feature importance, and range analysis visualizations, as well as a visualization highlighting typical model mistakes, and another showing data samples categorized by true/false positive/negative prediction scenarios. Collaboration was facilitated through a structured walkthrough where the researchers explained each visual explanation component, its meaning, and interactive features (such as help buttons for accessing scientific evidence). For each component, nurses were explicitly asked whether they would use it in practice and were asked to provide critiques and suggestions for improvement.  

The nurses' feedback led to substantial refinements in our approach, particularly in three critical areas: First, the nurses expressed a strong preference for individual explanations over global ones, noting that showing feature contributions for individual patients would align better with their natural diagnostic processes in clinical settings. This feedback prompted a shift from global model explanations to patient-specific analyses that aligned more closely with clinical workflows.

Second, the nurses emphasized the need for actionable recommendations that explained how patients at risk of Diabetes could reduce their risk. This led to the development of the step-based counterfactual explanations and the recommendation system to achieve the suggested changes in the health metrics.

Third, they provided critical feedback on several visualizations. For the visualization showing common model mistakes, nurses questioned its clinical relevance, despite being understandable. They expressed concerns that the classification visualization would be difficult for nurses to comprehend. In contrast, they found the Patient Record Visualization appropriate and useful for their clinical workflow.

Based on these insights, we redesigned our original conversational UI with in-chat visualizations to an integrated visual-conversational DSS that provides more intuitive access to key visualizations. We also incorporated a risk percentage display component to enhance clinical assessment capabilities. In the second co-design session, we presented the revised design featuring the dashboard-style interface with dedicated visualization panels alongside the conversational component. The nurses confirmed that the new design satisfied their expectations and better aligned with their clinical workflow needs, validating our redesign approach.

\section{User Study} \label{sec:user_study}

We conducted a mixed-methods user study with 30 healthcare professionals to evaluate the effectiveness of our integrated visual-conversational DSS for diabetes risk assessment. Below, we detail the study setup, evaluation measures, and study procedure.

\subsection{Study Setup}
The study was conducted online through a web-based platform, with participants taking an average of 55 minutes to complete the user study. Participants were recruited online through Prolific \cite{prolific} and compensated at an hourly rate of \$15.13. Table~\ref{tab:demographics} presents a summary of participant demographics and professional characteristics. The majority of participants were nurses (76.7\%), with representation from other healthcare roles including physicians, paramedics, and diabetes educators. The sample showed a strong female representation (83.3\%) and was predominantly middle-aged, with 43.3\% in the 30-40 age range. Notably, 70\% of participants had received specialized training in diabetes care management.

\begin{table*}[b]
\centering
\caption{Participant Demographics and Professional Characteristics}
\Description{Summary table of study participants (n=30) showing distribution across professional roles (predominantly nurses at 76.7\%), gender (mostly female at 83.3\%), age groups (majority in 30-40 range at 43.3\%), frequency of diabetes patient interaction, specialized training status, AI tool usage frequency, and geographic distribution across the UK, US, and Canada.}
\label{tab:demographics}
\renewcommand{\arraystretch}{1.2}

% First section with 4 columns
\begin{tabular}{p{3.5cm}|p{3.5cm}|p{3.5cm}|p{3.5cm}}
\hline
\rowcolor{gray!20} \textbf{Professional Role} & \textbf{Diabetes Patient\newline Interaction} & \textbf{Specialized Diabetes\newline Training} & \textbf{Geographic Distribution} \\
\hline
Nurse: 23 (76.7\%) & Daily: 13 (43.3\%) & Yes: 21 (70.0\%) & United Kingdom: 21 (70.0\%) \\
Physician: 2 (6.7\%) & Weekly: 14 (46.7\%) & No: 9 (30.0\%) & United States: 7 (23.3\%) \\
Paramedic: 2 (6.7\%) & Monthly: 3 (10.0\%) & & Canada: 2 (6.7\%) \\
Other Healthcare Roles: 3 (10.0\%) & & & \\
\hline
\end{tabular}

\vspace{0.3cm}

% Second section with 3 columns only
\begin{tabular}{p{3.5cm}|p{3.5cm}|p{3.5cm}}
\hline
\rowcolor{gray!20} \textbf{Gender} & \textbf{Age Group} & \textbf{AI Tool Usage} \\
\hline
Female: 25 (83.3\%) & 20-30: 7 (23.3\%) & Frequently (Weekly): 8 (26.7\%) \\
Male: 5 (16.7\%) & 30-40: 13 (43.3\%) & Occasionally (Monthly): 8 (26.7\%) \\
& 40-50: 8 (26.7\%) & Rarely (1-2 times/year): 7 (23.3\%) \\
% & 50-60: 1 (3.3\%) & Never: 7 (23.3\%) \\
& Above 60: 2 (6.7\%) & \\
\hline
\end{tabular}
\end{table*}

To avoid AI generated/assisted responses in our study, we implemented several validation measures. Responses were analyzed using Claude 3.5 Sonnet to detect patterns indicative of AI assistance in writing. Participants whose responses showed evidence of AI usage were excluded from the study. Additionally, participants who did not fully engage with the system, such as those who did not enter the system or asked significantly fewer questions or questions unrelated to our proposed query set, were classified as failing to adequately interact with the chatbot and excluded from the final analysis.

Similar to previous work that involved recruiting domain experts from crowd-sourcing platforms \cite{bhattacharya2024explanatory}, to ensure participants possessed adequate clinical expertise, we validated their domain knowledge through a comprehensive assessment. The evaluation covered several key areas of diabetes care expertise, including understanding of type 2 diabetes pathophysiology, diagnostic criteria, treatment protocols, complication management, and lifestyle modification guidance. This assessment helped ensure that participants had the necessary medical background to effectively evaluate the system's clinical utility. 

\subsection{Study Procedure}
Prior to beginning the study tasks, participants were introduced to the different components of the DSS through a video tutorial that demonstrated the system's key features and functionality. The study then proceeded with instructing the participants to interact with the system through four tasks, each designed to evaluate specific aspects of the system.

In Task 1 (Initial Assessment and Monitoring), participants were asked to examine two patient cases using the patient record and risk assessment visualizations. Specifically, they were instructed to enter two patient IDs into the system, which then displayed the corresponding patient information (such as the patient with ID 39 shown in Figure \ref{fig:system_interface}). They first assessed these cases without using the chatbot to determine which patient's condition was more critical and explain their reasoning. They then used the chatbot to ask follow-up questions about the risk assessment for a specific patient, exploring how the conversational interface helped their understanding beyond the visualizations.

Task 2 was divided into two parts focusing on understanding the system's analytical capabilities. In Task 2A (Feature Importance Visualization), participants identified the top contributing factors to the prediction and examined the scientific evidence provided for factor importance by accessing help documentation and citations. They were encouraged to use the chatbot to ask comparative questions about factor importance. Task 2B (Feature Range Visualization) required participants to explore the relationship between AI-observed and scientific ranges, including examining the overlap between these ranges and accessing supporting scientific evidence through help icons.

In Task 3 (Action Planning Through Recommendations), participants worked with the recommendation visualization to examine risk reduction steps. They used the "Get Recommendations" feature to access detailed recommendations and engaged with the chatbot to obtain more specific guidance about implementing these suggestions. This task focused on evaluating how well the system supported actionable intervention planning.

Task 4 (Evaluating Chatbot Interaction) involved interaction with the chatbot across all visualization types. Participants were prompted to ask specific questions about various aspects of the system, including queries about factor inclusion, range alignment, and implementation timelines for recommendations. This task was designed to assess how well the conversational interface supported deeper understanding of the system's assessments.

For each task, participants provided both quantitative and qualitative feedback. The quantitative evaluation used Likert-scale ratings across four key dimensions: Explanations Satisfaction, Understandability, Actionability, and Trust. Qualitative feedback was collected through open-ended questions that explored participants' experiences with specific system features and their suggestions for improvement. Throughout the study, the system recorded detailed interaction data, including visualization usage patterns and chat conversation logs, to provide objective measures of engagement with different system components.

Our task design methodology followed a systematic approach grounded in prior research. The evaluation metrics and the core tasks (Tasks 1, 2A, and 3) were adapted from the evaluation framework by Bhattacharya et al. \cite{bhattacharya2023directive}, which demonstrated effectiveness in assessing healthcare visualization systems. For the additional components of our system, specifically the conversational interface and feature range explanations, we designed additional complementary tasks (Tasks 2B and 4) using similar structural principles but targeting these new interaction modalities. 

\subsection{Evaluation Measures}
Our evaluation framework drew upon established metrics from multiple sources in the explainable AI literature. For Tasks 1, 2, and 3, the Likert-scale questions measuring Explanations Satisfaction, Understandability, and Trust were adapted from the evaluation framework proposed by \cite{hoffman2018metrics}. The Actionability questions for these tasks were derived from \cite{singh2024actionability}. For Task 4, which focused on the conversational interface, we adapted the Likert-scale questions from the XEQ Scale \cite{wijekoon2024xeq}. In all cases, the questions were modified to align with our specific application context of diabetes risk assessment and clinical decision support.

The open-ended questions were designed to align with our research questions and were inspired by the XAI question bank presented in \cite{liao2020questioning}. This approach allowed us to explore how healthcare professionals understood and interacted with the system's explanations, how scientific evidence influenced their trust, and how the integrated visual-conversational approach supported their clinical decision-making processes. All evaluation measures were carefully adapted to ensure relevance to the healthcare context while maintaining the validated structure of the original assessment frameworks.

\subsection{Thematic Analysis}

To analyze the qualitative data collected from the open-ended questions, we employed a hybrid approach of deductive and inductive coding in our thematic analysis  \cite{clarke2017thematic, fereday2006demonstrating}. This methodological choice combined theory-driven deductive coding with data-driven inductive coding, allowing us to integrate established theoretical frameworks while remaining responsive to novel patterns emerging from the data. The analysis process began with the development of an initial set of deductive codes informed by previous research findings \cite{bhattacharya2023directive, rajashekar2024human}. The codes were derived by the principal researcher prior to the formal coding phase to specifically address the research questions concerning conversational explanations' impact on understandability, scientific evidence's role in trust calibration, and the usefulness of integrated visual-conversational explanations. From this process, we established 13 deductive codes that guided our initial analysis.

The coding procedure followed a structured approach to ensure analytical rigor through inter-coder reliability analysis. First, two researchers thoroughly familiarized themselves with the data through multiple readings of participant responses. Then, both coders independently analyzed one-third of the data using the deductive codes. After this initial coding phase, the researchers compared their coding decisions, discussed discrepancies, and clarified code interpretations to establish a shared understanding before proceeding to code the remaining data independently \cite{o2020intercoder}. Following the work of \cite{fereday2006demonstrating} for our thematic analysis, the reliability assessment focused specifically on the deductive codes. To quantify the level of agreement between coders, we calculated Cohen's kappa across these deductive codes. The analysis yielded a kappa value of 0.847, indicating substantial agreement according to standard interpretations of this metric \cite{landis1977measurement}. This high level of agreement provided confidence in the reliability of our coding framework and the resulting thematic analysis.

In parallel with the deductive coding process, the principal researcher conducted a separate round of inductive coding to identify emerging patterns not captured by the predetermined deductive codes. This process generated 24 inductive codes, of which 13 were ultimately incorporated into our final thematic structure. During the synthesis phase, these inductive codes were used in combination with the deductive codes to develop comprehensive themes and sub-themes that represented participants' perspectives.

After coding was complete, we synthesized the coded data into coherent themes and sub-themes through an iterative process of pattern recognition and refinement. The final thematic structure was organized according to our three research questions, with each theme supported by representative quotes from participants and an indication of how widely each theme was represented across the sample. Table~\ref{tab:thematic_analysis} provides an overview of the main themes and sub-themes organized by research question, indicating whether they are derived from a deductive code (D), an inductive code, or a combination of both codes (D-I).

\begin{table*}[!b]
\caption{Overview of Qualitative Findings with Supporting Participants}
\label{tab:thematic_analysis}
\renewcommand{\arraystretch}{1.6} % Add space between rows and lines
\begin{tabular}{p{0.6\textwidth}|p{0.3\textwidth}}
\hline
\rowcolor{gray!20} \textbf{Themes and \textit{Sub-themes}} & \textbf{Supporting Participants\newline (n/30): IDs} \\
\hline
\multicolumn{2}{c}{\vspace{0.2em}\textbf{\large RQ1: Impact of Conversational Explanations on Understandability}\vspace{0.2em}} \\
\hline
Interactions with the chatbot increase interpretability 
of the explanation methods (D) & (29/30): 1-8, 10-30 \\
% \hline
\textit{Textual explanations clarified visuals, but after a while,
a chatbot may become unnecessary} (D-I) & (20/30): 1, 3-5, 7-10, 13-14, 
16, 18-21, 23-24, 26, 29-30 \\
\hline
HCPs value textual information that is easy to understand and quick to read (D-I) & (16/30): 1-4, 8, 10, 12, 15, 
19-20, 22-24, 27-28 \\
\hline
\multicolumn{2}{c}{\vspace{0.2em}\textbf{\large RQ2: Impact of Scientific Evidence on Trust Calibration}\vspace{0.2em}} \\
\hline
Evidence-based explanation builds trust in AI (D) & (28/30): 1-3, 5-8, 10-30 \\
\hline
AI cannot fully replace a clinician and therefore cannot be trusted fully (D) & (8/30): 2, 5, 9, 14, 17, 20-21, 26 \\
\hline
Uncertainty about data sources leads to distrust of AI output (D) & (8/30): 3, 5, 7, 10, 18, 20, 25, 28 \\
\hline
Lack of understanding of a visualization causes mistrust of it (D) & (2/30): 4, 17 \\
\hline
\multicolumn{2}{c}{\vspace{0.2em}\textbf{\large RQ3: Usefulness of Integrated Visual and Conversational Explanations}\vspace{0.2em}} \\
\hline
HCPs used both visual and conversational components to analyze risk and get recommendations (D) & (25/30): 1-3, 5-8, 11-13, 
15-16, 18-27, 29 \\
\hline
Dashboard combines patient data for faster risk analysis, but the information remains insufficient (D-I) & (24/30): 1-10, 14-17, 19, 
21-23, 26, 28, 30 \\
\textit{Visualizations facilitate quick
understanding} (D) & (7/30): 1, 5, 14, 16, 19, 22, 28 \\
\hline
Recommendations are too generic and need to be personalized based on comprehensive patient background (D) & (26/30): 1-5, 8, 10-11, 13-21, 
23-24, 26-27, 29-30 \\
\hline
\end{tabular}
\end{table*}

\section{Findings} \label{sec:findings}

In this section, we present the findings from our mixed-methods study with healthcare professionals. The results are organized according to our three research questions, integrating both qualitative and quantitative data to provide a comprehensive understanding of healthcare professionals' experiences with our system. For each qualitative finding, we indicate its source and whether it was identified deductively (D), inductively (I), or both (D-I).

Throughout the analysis, we examine users' interactions with the system, which showed active engagement during the average 41-minute session duration per participant. The chat interface demonstrated particularly high engagement, with an average of 17.32 chat interactions per user compared to 5.32 visualization interactions, suggesting that participants actively utilized the conversational capabilities for deeper understanding. Figure~\ref{fig:rating_distributions} presents the mean ratings and standard deviations across all tasks and dimensions.

\subsection{RQ1: Impact of Conversational Explanations on Understandability}

\subsubsection{Interactions with the chatbot increase interpretability of the explanation methods (D) \cite{bhattacharya2023directive}}

The majority of participants (29/30) found that interactions with the chatbot enhanced their ability to explore and understand the system's explanations. The conversation allowed healthcare professionals to receive reasoning and supplemental information that clarified the AI's decision-making process.
As P29 described:
\begin{quote}
``[The most helpful aspect of conversing with the AI was the] ability to interact with it to gain further insights and clarification on the data provided. Being able to ask follow-up questions on things like how certain factors influence risk or why certain factors were included helped me better understand how the AI was processing the data.''
\end{quote}

\begin{figure*}[t!]
  \includegraphics[width=\textwidth]{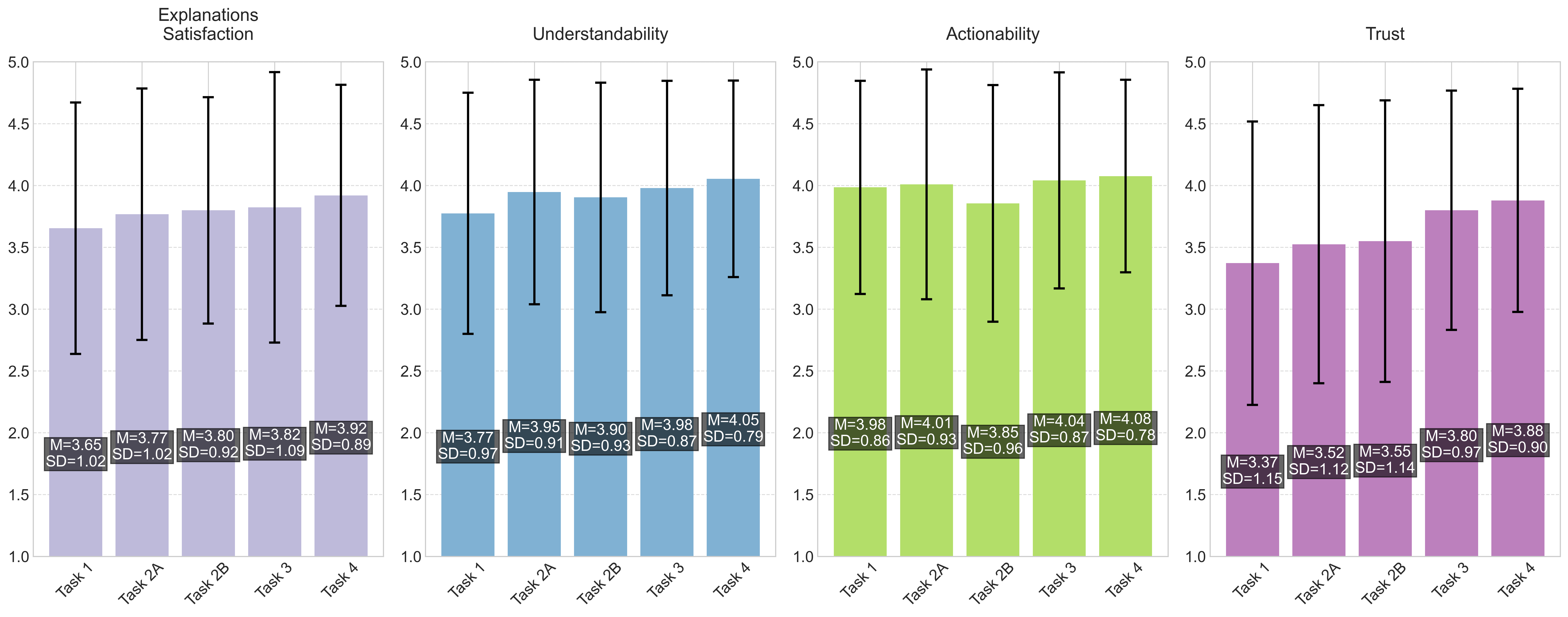}
  \caption{Mean and standard deviations of participants' ratings across all tasks}
  \Description{A chart showing the mean ratings and standard deviations of participants' responses across all evaluation tasks, measuring four dimensions: Explanations Satisfaction, Understandability, Actionability, and Trust.}
  \label{fig:rating_distributions}
\end{figure*}

This qualitative finding is supported by our quantitative results, as shown in Figure \ref{fig:rating_distributions}, where Task 4 (Chatbot Interaction) received the highest Understandability ratings ($M = 4.05$, $SD = 0.79$) across all tasks. Also, the chatbot interaction showed an improvement in understandability compared to the initial assessment task (Task 1, $M = 3.77$, $SD = 0.97$), suggesting that the conversational component supported healthcare professionals' understanding of the system.

\begin{figure*}[t!]
    \centering
    \begin{subfigure}[b]{0.4\textwidth}
        \centering
        \includegraphics[width=\textwidth]{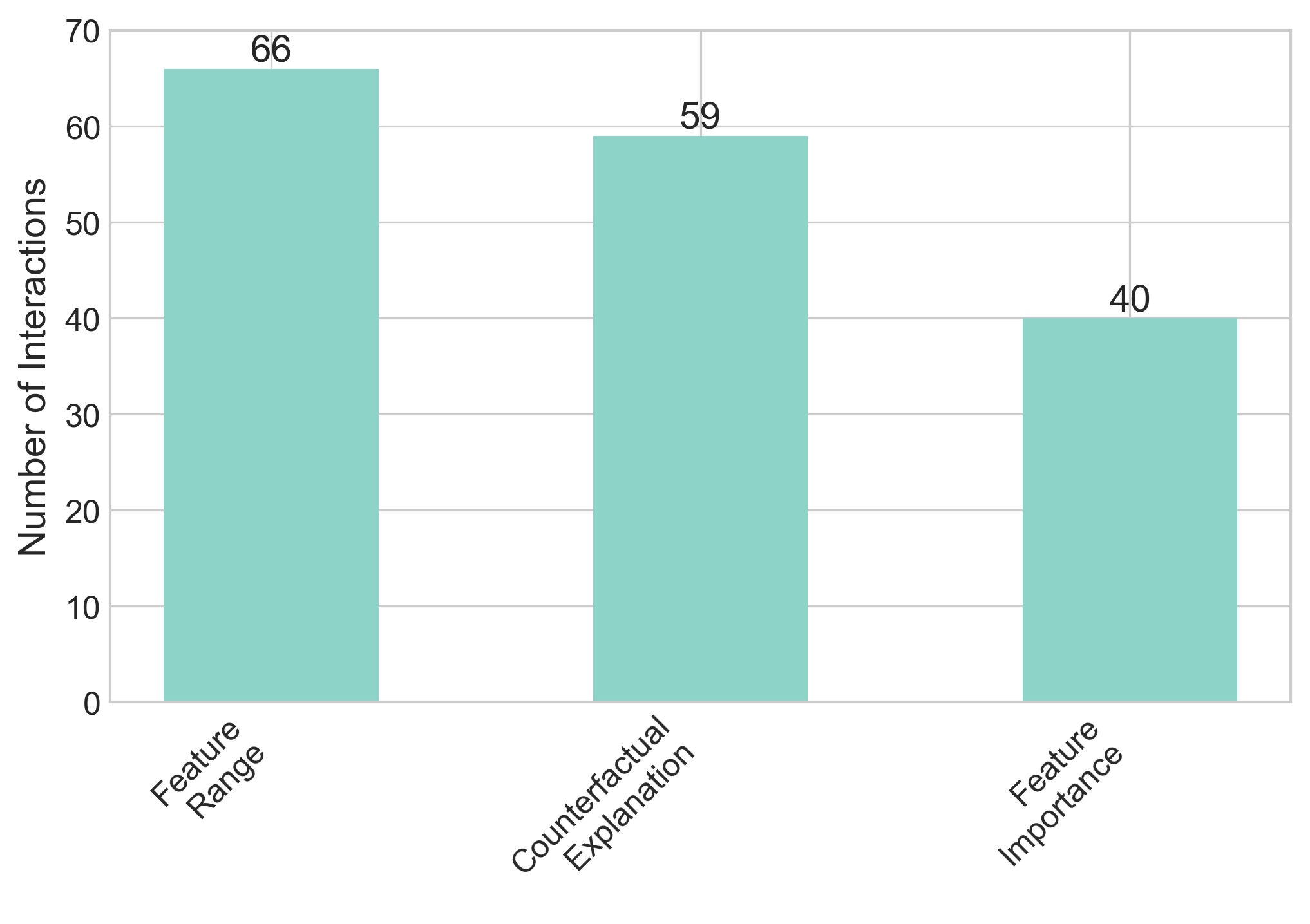}
        \caption{Distribution of visualization interactions}
        \label{fig:visualization_interactions}
    \end{subfigure}
    \hfill
    \begin{subfigure}[b]{0.52\textwidth}
        \centering
        \includegraphics[width=\textwidth]{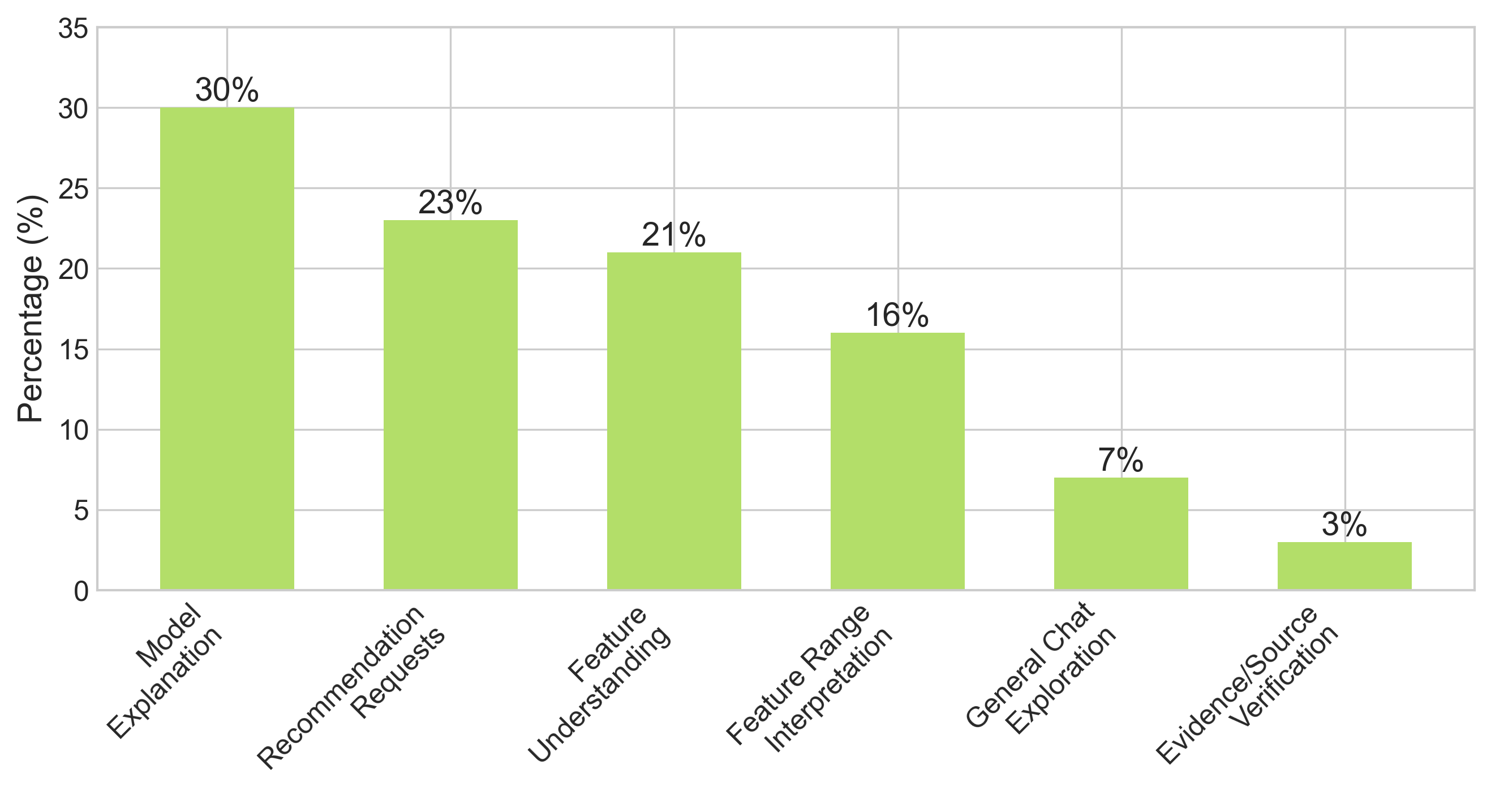}
        \caption{Distribution of question types}
        \label{fig:question_distribution}
    \end{subfigure}
    \caption{Analysis of user interactions: (a) shows the frequency of different visualization types used, with Feature Range Analysis being the most utilized component; (b) illustrates the distribution of different types of questions asked by users, with Model Explanation queries being the most common.}
    \label{fig:interaction_analysis}
    \Description{Two charts analyzing user interaction patterns: (a) shows the frequency distribution of different visualization types used, with Feature Range Analysis being the most frequently utilized component; (b) illustrates the distribution of question types asked by users, with Model Explanation queries being the most common category.}
\end{figure*}

Analysis of chat interactions revealed that model explanation queries formed the largest category (30\% of all questions), with practitioners seeking to understand the AI's decision-making process through questions like ``Can you explain why this patient has been flagged as high risk for diabetes?'' as illustrated in Figure~\ref{fig:question_distribution}. Feature understanding questions constituted another 21\% of interactions, with healthcare professionals investigating the relative importance of different factors. The substantial proportion of these question types (51\% combined) indicates that healthcare professionals actively engaged in building a comprehensive understanding of the AI's reasoning process.

Within this theme, we identified a notable sub-theme related to the complementary nature of textual and visual explanations.

\paragraph{Textual explanations clarified visuals, but after a while, a chatbot may become unnecessary (D-I)}
Many participants (20/30) indicated that textual explanations provided by the chatbot helped clarify the visualizations, particularly when they encountered difficulty interpreting certain visual elements. This was especially evident with the feature range explanation, where several participants (8/30) noted that the chatbot helped them understand the difference between AI-observed ranges and scientific ranges. As P8 stated:
\begin{quote}
``It was very helpful [to ask about feature ranges] because I didn't exactly understand what the difference was between the AI-observed range and the scientific range. Now I understand the key differences between both.''
\end{quote}

This is reflected in our interaction data, where Feature Range Analysis received the highest engagement (66 interactions) among visualization components, followed by Counterfactual Explanation (59 interactions) and Feature Importance (40 interactions), as shown in Figure~\ref{fig:visualization_interactions}. Some participants (2/30) suggested that the chatbot might become unnecessary as healthcare professionals become more familiar with the visualizations, with P19 noting that ``after a period of using the data and visualization alone, it is possible that the clinician will not use the chatbot so frequently.''

\subsubsection{HCPs value textual information that is easy to understand and quick to read (D-I) \cite{rajashekar2024human}}

Many participants (16/30) emphasized the value of receiving information in an intuitive way that can be read quickly. They valued the AI's ability to provide reasoning in a comprehensible way (9/30) and appreciated that the chatbot's responses do not take too long to read (7/30). As P12 expressed:
\begin{quote}
``[The conversation with the AI] was quicker than when I Google something and gives me the facts without me needing to read through pages and pages of information to find what I want.''
\end{quote}

Some healthcare professionals (3/30) specifically noted the AI assistant's ability to summarize data into an understandable format:
\begin{quote}
``The AI was able to give a summary of the data in a simple to read and understand format whereas I would have to read through the data and come to a conclusion myself. The AI doing it for me helped in this process.'' (P8)
\end{quote}

\subsection{RQ2: Impact of Scientific Evidence on Trust Calibration}

\subsubsection{Evidence-based explanation builds trust in AI (D) \cite{bhattacharya2023directive, rajashekar2024human}}

The inclusion of scientific evidence in AI explanations positively influenced trust for nearly all participants (28/30). Healthcare professionals valued the system's grounding in evidence-based practice, which aligns with their professional standards. As P19 emphasized:
\begin{quote}
``This [evidence] is absolutely necessary when a clinician is interacting with an AI system. When evidence and reference is provided, not only it means that the AI decision is evidence based but also it shows with good confidence that it is not generated by the AI system in error.''
\end{quote}

The quantitative data supports this finding, as shown in Figure \ref{fig:rating_distributions}, showing a progressive increase in Trust ratings across tasks, from the initial assessment (Task 1, $M = 3.37$, $SD = 1.15$) through to chatbot interaction (Task 4, $M = 3.88$, $SD = 0.90$). Many participants (11/30) specifically appreciated that the scientific evidence came from reputable sources they were already familiar with from their own research.

The overlap between AI-observed ranges and scientific ranges had a particularly positive effect on building trust for many participants (17/30):
\begin{quote}
``If the difference [between AI observed and scientific ranges] was high then I would be more skeptical of the AI as we know that scientific data is robust and has been analyzed thoroughly however with the overlap shown it enhances my confidence as it shows that the AI is aligned with previous scientific research.'' (P22)
\end{quote}
This is reflected in the quantitative results (Figure \ref{fig:rating_distributions}) for Task 2B (Range Analysis), which showed improved Trust ratings ($M = 3.55$, $SD = 1.14$) compared to Task 1.

\subsubsection{AI cannot fully replace a clinician and therefore cannot be trusted fully (D) \cite{rajashekar2024human}}

Despite the positive impact of scientific evidence, several participants (8/30) expressed the belief that AI could not fully replace clinical judgment and, therefore, could not be trusted completely. As P17 noted, they would always seek a second opinion because they ``don't want to entrust someone's health fully with a computer.'' Others indicated a desire to verify the cited sources themselves rather than accepting them uncritically.

This healthy skepticism led some healthcare professionals to want to verify the cited sources themselves rather than accepting them uncritically: \begin{quote}
``It [scientific evidence] slightly increased trust, but just because a paper is 	referenced does not mean that it is scientifically robust or clinically significant, I would want to review 	the source myself to see if it was trustworthy.'' (P2)
\end{quote}

\subsubsection{Uncertainty about data sources leads to distrust of AI output (D) \cite{rajashekar2024human}}

Some participants (8/30) expressed concerns about trusting the AI's outputs due to unclear information about the underlying data sources. Participants noted that while they had confidence in the scientific ranges, they lacked knowledge about the datasets used by the AI. As P18 stated:
\begin{quote}
``I have a lot of confidence in the scientific ranges and I don't know enough about the datasets the AI is using to know how accurate its assessment is.''
\end{quote}

These finding helps explain why Trust ratings, while showing improvement across tasks, remained lower than other dimensions such as Understandability and Actionability.

A smaller number of participants (2/30) worried about the AI's ability to remain current with the latest medical research. This finding may partially explain the higher variability observed in Trust ratings (SDs ranging from 0.90 to 1.15) compared to other dimensions shown in Figure \ref{fig:rating_distributions}.

\subsubsection{Lack of understanding of feature range visualization causes mistrust in AI (D) \cite{bhattacharya2023directive}}

A small number of participants (2/30) indicated that difficulty understanding the feature range visualization specifically led to a lack of trust in AI's assessments. For instance, P17 found the ranges ``quite confusing'' and noted that the variations made it difficult to trust the AI.

\subsection{RQ3: Usefulness of Integrated Visual and Conversational Explanations}

\subsubsection{HCPs used both visual and conversational components to analyze risk and get recommendations (D) \cite{bhattacharya2023directive}}

Most participants (25/30) found value in having both visual and conversational modalities available for risk analysis. As P3 noted, the system ``displays metrics in an easy to use interface'' while showing ``the risk and how that risk has been determined when questioned.'' The integration of both modalities was particularly valued for developing recommendations, with P22 highlighting that ``the combination of both allows for a robust plan to be put in place to help reduce risk.''

This finding is supported by the quantitative data, shown in Figure \ref{fig:rating_distributions}, for Task 3 (Recommendation System), which showed improvements across all dimensions compared to previous tasks, with particularly high ratings for Actionability ($M = 4.04$, $SD = 0.87$) and Understandability ($M = 3.98$, $SD = 0.87$), as shown in Figure~\ref{fig:rating_distributions}. 

Several participants (5/30) specifically appreciated how the system presented manageable steps to reduce risk, while others (4/30) valued the ability to ask follow-up questions about recommendations. Analysis of chat interactions revealed that recommendation requests constituted 23\% of all questions, indicating that participants valued the system's potential for practical clinical application.

\subsubsection{The dashboard combines patient data for faster risk analysis, but the information remains insufficient (D-I) \cite{bhattacharya2023directive}}

Most participants (24/30) appreciated how the dashboard consolidated patient information, making risk analysis faster and more efficient. As P15 expressed:
\begin{quote}
``It gives me some key metrics to perform an initial risk status. Clear. I don't need to scrawl through the patient record to find what I need.''
\end{quote}

This finding aligns with the quantitative results (Figure \ref{fig:rating_distributions}) for Task 1 (Initial Assessment and Monitoring), which showed Actionability ratings above the scale midpoint ($M = 3.98$, $SD = 0.86$). However, several participants (4/30) indicated that the presented information was insufficient for comprehensive risk analysis, expressing the need for additional health metrics such as HbA1c, lipid levels, or genomic information.

\paragraph{Visualizations facilitate quick understanding (D) \cite{bhattacharya2023directive}}
Several participants (7/30) specifically noted that the visual representations enabled quick comprehension of the explanations. P22 observed that the dashboard ``provides a quick and concise overview of factors contributing to the individual's risk of diabetes'' and ``allows the practitioner to clearly see what areas are concerning and what areas are acceptable.''

\subsubsection{Recommendations are too generic and need to be personalized based on comprehensive patient background (D) \cite{bhattacharya2023directive}}

Most participants (26/30) felt that the system's recommendations were too generic and required more personalization. As P4 noted, the recommendations ``somewhat help, but not as specific as it needs the be'' since a plan needs to be developed with the patient that is ``specific to their particular needs.''

Many healthcare professionals emphasized the importance of considering individual factors when developing recommendations. P10 stressed that psychological factors, barriers to change, and economic barriers ``HAVE to be taken into consideration when making recommendations'' because ``one size fits all approaches DO NOT WORK.''

Despite these criticisms, Task 3 (Recommendation System) received high Actionability ratings ($M = 4.04$, $SD = 0.87$), suggesting that participants still found value in the recommendations as a starting point for patient care. The 23\% of chat interactions focused on recommendation requests further highlights the importance participants placed on actionable guidance, despite their desire for more personalized recommendations. This suggests that while the current implementation provides value, there is potential for improvement in tailoring recommendations to individual patient circumstances.

\section{Discussion} \label{sec:discussison}

Our study explored the integration of conversational AI capabilities with interactive visualizations in a clinical decision support system for diabetes risk assessment. Through a mixed-methods evaluation with 30 healthcare professionals, we examined how this integration impacts understandability, trust calibration, and clinical utility. This section discusses our findings in relation to prior research and their implications for the design of AI-powered clinical decision support systems.

\subsection{The Role of Conversational Explanations in Understanding}

Our findings demonstrate that conversational explanations enhance healthcare professionals' understanding of AI-driven clinical assessments. Nearly all participants (29/30) reported that interactions with the chatbot improved their ability to explore and understand the system's explanations, with Task 4 (Chatbot Interaction) receiving the highest understandability ratings across all tasks.

These findings extend prior research by \citet{bhattacharya2023directive} who observed that "including textual annotations can improve understandability" of visualizations and \citet{rajashekar2024human} who noted that "participants found the graphical representation challenging to interpret" and that "large language models may increase ease of use for AI-CDSS." While their participants expressed difficulty interpreting data-distribution charts without sufficient textual context, our participants particularly valued the chatbot for clarifying complex visualizations, especially the feature range analysis, which received the highest engagement among visualization components. Moreover, our approach extends \citeauthor{bhattacharya2023directive}'s observation that "interactive visuals increase the interpretability of explanation methods" \cite{bhattacharya2023directive} by demonstrating how conversational explanations further support understandability through targeted follow-up questions to gain more insights into the explanations of the AI's decision-making.

Importantly, our findings align with \citet{rajashekar2024human} and build upon \citet{Slack2023} regarding the benefits of conversational explanations for healthcare professionals with limited technical AI knowledge. Healthcare professionals value textual information that is easy to understand and quick to use. This preference for efficient information delivery reflects the time constraints of clinical environments, where practitioners need to process information rapidly. \citet{Slack2023} noted a distinct difference in how technical expertise influences interface preferences—healthcare workers showed greater future use intention for conversational interfaces compared to ML professionals with technical backgrounds. This connects with our observation that some participants suggested the chatbot might become less necessary as they become more familiar with interpreting the visualizations. This pattern indicates that conversational explanations may serve as a valuable bridge to visual literacy, with their relative importance potentially decreasing as users develop expertise in interpreting visual representations of AI decisions.

\subsection{Evidence-Based Explanations and Trust Calibration}

Our research highlights the critical role of scientific evidence in calibrating healthcare professionals' trust in AI systems. The incorporation of evidence-based explanations positively influenced trust for nearly all participants (29/30), with progressive improvements in trust ratings observed across tasks as exposure to evidence-grounded explanations increased.

Our explicit integration of scientific evidence alongside AI explanations proved crucial for trust calibration, particularly when the evidence came from reputable sources familiar to participants. This aligns with \citeauthor{rajashekar2024human}'s finding that "large language models require justification with citations to promote trust" and \citeauthor{yang2023harnessing}'s assertion that scientific evidence must be "drawn from a shared source of truth" with healthcare professionals.

However, our findings reveal a more nuanced relationship between AI explanations and scientific evidence than previously suggested. While \citet{yang2023harnessing} proposed that scientific evidence should largely supplant explanations of AI decision-making processes, our study indicates these approaches should be complementary. Several participants (8/30) expressed uncertainty about the AI's underlying data sources, which contributed to distrust of its outputs. This is consistent with \citet{rajashekar2024human} who found that "the most commonly cited reason for why participants could not trust the chatbot's outputs was that they did not know what data the chatbot was drawing from."

The feature range analysis visualization, which explicitly compared AI-observed ranges with scientific ranges, proved particularly valuable for trust calibration. Many participants (17/30) noted that the overlap between these ranges enhanced their confidence in the system. This suggests that transparency in both the AI's decision-making process and its alignment with established medical knowledge is essential for building appropriate trust.

Our findings extend Dazeley et al.'s argument that explaining a single decision without the origins of that action or its context will not carry sufficient meaning \cite{dazeley2021levels}. By providing scientific evidence as context for AI explanations, our system creates a framework that allows healthcare professionals to evaluate AI decisions against established medical knowledge. our study revealed a healthy skepticism among healthcare professionals (8/30) who maintained that AI cannot fully replace clinical judgment and emphasized the need to verify sources independently.
This suggests that evidence should not simply be presented as a persuasive tool but rather as material for critical evaluation.

\subsection{Complementary Nature of Visual and Conversational Modalities}

Our findings demonstrate that healthcare professionals value having both visual and conversational modalities available for risk analysis and recommendation development. Most participants (25/30) reported using both components to assess patient risk and develop recommendations, with Task 3 (Recommendation System) showing high ratings for actionability ($M = 4.04$) and understandability ($M = 3.98$).

This preference for integrated modalities aligns with \citet{rajashekar2024human} who similarly found that clinicians appreciated having multiple interaction options. While they focused on contrasting interactive dashboards with chatbot interfaces, our study demonstrates how these modalities can be synergistically combined to support different aspects of clinical decision-making. The integration supports both analytical reasoning (through visualizations) and contextual understanding (through dialogue), enhancing the system's overall utility.

Consistent with \citet{bhattacharya2023directive} who found that "visualizations facilitate quick understanding" through graphical representations and color-coding, our participants valued visual components for rapid comprehension. Several participants (7/30) specifically noted that visualizations enabled them to "clearly see what areas are concerning and what areas are acceptable." Simultaneously, the conversational interface provided deeper insights and clarifications when needed, particularly for complex visualizations like the feature range analysis.

However, our study also identified limitations in the current implementation. Most participants (24/30) noted that while the dashboard effectively consolidated patient information for faster risk analysis, the presented data remained insufficient for comprehensive assessment. Similarly, many participants (26/30) found the recommendations too generic and lacking personalization based on patient-specific factors such as comorbidities, socioeconomic status, and psychological barriers to change.

These limitations mirror \citeauthor{bhattacharya2023directive}'s findings that healthcare professionals preferred data-centric explanations over counterfactual recommendations, considering the latter too generic and insufficiently personalized. However, our conversational interface offered some advantages for actionability by allowing healthcare professionals to ask follow-up questions about specific recommendations, with recommendation requests constituting 23\% of all chat interactions.

\subsection{Limitations}

Despite implementing assessments to validate domain knowledge of the participants, our study has several limitations. The primary limitation stems from the lack of formal verification of professional backgrounds beyond self-reporting through the Prolific platform. Our participant sample (83.3\% female) may not fully represent all healthcare practitioners. Additionally, the Pima Indians Diabetes Database, while widely used in research, is relatively small (768 samples) and lacks demographic diversity, potentially limiting generalizability. The moderate accuracy (73.3\%) of our machine learning model may have influenced healthcare professionals' trust perceptions and interaction patterns.

The choice of Claude Haiku as our conversational AI component, while effective for our study, may not represent the most advanced reasoning capabilities currently available. A design limitation of our approach is that we did not provide comprehensive evidence in terms of ``supporting and opposing evidence'' that healthcare professionals need to make an informed judgment \cite{yang2023harnessing}. Our system primarily presented evidence supporting the importance of various factors and their ranges rather than offering a balanced view of conflicting research findings, which would be necessary for more nuanced clinical decision-making.

\subsection{Implications for Future Research}

Our findings suggest directions for future research of integrated visual-conversational systems for healthcare decision support:

\paragraph{Adaptive Conversational Interfaces Based on User Expertise}
Our finding that healthcare professionals might rely less on conversational assistance as they become familiar with visualizations suggests developing explanation systems that adapt to users' growing expertise. Future systems could adjust both the extent of conversational support and the technical depth of explanations based on detected user sophistication or explicit preferences. For novice users, systems would abstract away AI complexities, focusing on clinically relevant interpretations using familiar medical terminology. As users demonstrate increased understanding through more sophisticated queries or interaction patterns, the system could progressively introduce technical concepts about feature interactions, algorithmic confidence, or statistical foundations. Research should explore methods for modeling user expertise through interaction analysis, creating seamless transitions between explanation modes, and designing interfaces that support different learning trajectories while maintaining clinical relevance.

\paragraph{Transparency in Data Sources and Model Development}
While scientific evidence improves trust in AI explanations, our findings show this alone is insufficient for complete trust calibration. Healthcare professionals need greater transparency about data sources used by both predictive models and explanatory systems. Current approaches, including ours, do not incorporate medical literature directly into model training. Instead, models make predictions based solely on patient data, with medical literature serving as external validation presented alongside outputs. This evidence exists outside the model and supports or opposes predictions without providing causal explanations of how scientific knowledge influenced specific predictions. Future work should explore training predictive models that incorporate both patient data and medical guidelines, then provide intuitive, efficient causal explanations by identifying which medical literature influenced risk assessments. This integration would bridge the gap between correlation-based machine learning and evidence-based clinical practice, creating AI systems that explain their reasoning using the same scientific evidence that clinicians rely upon.

\paragraph{Personalized, Context-Aware Recommendations}
Although participants appreciated the step-by-step approach to risk reduction, they found recommendations too generic for clinical practice. Future systems need to incorporate psychological factors, including patient preferences, time constraints, and socioeconomic status, that influence treatment adherence. The ability to ask follow-up questions about recommendations was valued for obtaining more relevant information, but this feature requires enhancement to address the full complexity of patient circumstances. Research should focus on developing recommendation frameworks that balance evidence-based guidelines with personalization capabilities, potentially incorporating shared decision-making approaches that involve both clinician and patient perspectives.

\section{Conclusion}
This paper investigated the integration of interactive visualizations with conversational AI for explaining diabetes risk predictions, focusing on three main areas: the impact of conversational explanations on understanding AI decisions, the role of scientific evidence in trust calibration, and the effectiveness of combining visual and conversational explanations. We presented a visualization-chatbot DSS for diabetes prediction, a methodology for grounding AI explanations in scientific evidence, a hybrid approach for handling diverse user prompts, a feature range analysis technique for deeper insight into AI decision-making, and empirical insights into healthcare professionals' interaction with AI explanations. Through a mixed-methods study with 30 healthcare professionals, we demonstrated that conversational interaction supported comprehension of AI assessments, while integration of scientific evidence led to appropriate trust development. The combined visual-conversational approach proved effective in helping both risk evaluation and recommendation development.

\begin{acks}
This research is funded by the Research Foundation–Flanders (FWO grant G0A4923N). We sincerely thank Yizhe Zhang and Maxwell Szymanski for their insightful feedback.
\end{acks}

\bibliographystyle{ACM-Reference-Format}
\bibliography{sample-base}

% \appendix

\end{document}